# A Study of Obstacles in Plagiarism Software Subscribing by Colleges in Tamil Nadu

Subaveerapandiyan A.[#,*] and Sakthivel N.[$]

[#]DMI - St. Eugene University, Lusaka, Zambia
[$]Jio Institute Digital Library, Jio Institute, Mumbai - 410 206, India
[*]E-mail: subaveerapandiyan@gmail.com

## ABSTRACT

This article attempts to comprehend the current issues and hurdles that Indian colleges affiliated with Tamil Nadu State Universities encounter when trying to subscribe to a software that detects plagiarism. The study's goals are to determine whether colleges employ anti-plagiarism software, whether they ensure that their student given assignments are free of copyright infringement, whether tutors teach about academic misconduct, and what people seem to think of anti-plagiarism software. We surveyed for this study and distributed the questionnaires among college administrators, principals, and librarians. The study respondents are 15.9 per cent principals, 64.2 per cent library professionals, and 19.9 per cent college administrators. The survey study report shows that 70.9 per cent of the majority of the colleges did not subscribe. 41.7 per cent gave the reason it is too expensive, and 30.5 per cent of respondents thought that for their college, it is unnecessary to subscribe. However, nobody has confirmed whether or not all colleges possess access to plagiarism detection software. Thus, according to this investigation, further Indian states must be involved in this research to understand the specific context fully. This report advises the UGC to enforce the requirement that colleges have plagiarism detection software; they either provide colleges additional money to subscribe to such software, or the university must grant free access to the affiliated colleges.

## 1. INTRODUCTION

Academic offense, sometimes known as academic misconduct, is a severe problem in the twenty-first century. Plagiarism in research indicates malfeasance. Representing another's intellectual output to be one's own is morally and ethically unacceptable. It uses a word, phrase, or concept without properly acknowledging its source (s) or author(s). The most prevalent types of plagiarism are: (i) self-plagiarism, also characterised as replication of one's writing, is the utilisation of one's own formerly published materials, ideas, and principles. Even without the publisher's knowledge, authors occasionally notify publishers to copyright their published works, a practice known as self-plagiarism[1]., (ii) Translation plagiarism (or cross-language plagiarism) is challenging to spot because it was plagiarised from a different regional vernacular without crediting the respective writers[2]., (iii) Intentional plagiarism is when someone deliberately copies another person's work and uses ideas or phrases from that person's work without giving credit to them, (iv). Unintentional plagiarism occurs when the author(s) does not know how to cite the actual source properly, is indeed not aware of the citation rules and the plagiarism guideline, and neglects to adhere to them[3]., and (v) Mosaic/patch writing is plagiarism in which the text is reproduced, some phrases or words used modified, and the text isn't enclosed in quotation marks. It's possible that it was plagiarised through one or even more sources[4].

Twenty-two state universities and more than 3251 affiliated institutions are situated in Tamil Nadu[5]. These 22 universities provide a broad range of courses and specialisations. The general category (Multidisciplinary) includes nine universities. Thirteen universities are made up of the following disciplines: (Technology, Women Only, Tamil, Agricultural, Legal, Medical, Fishery, Fine Arts, Law, Distance Education, Sports, Physical Education, Teacher Education, and Veterinary Science)[6].

### 1.1 Importance of Scholarly Writing and Publishing

Writing is a kind of art. Academic writing is yet another word for scholarly writing. Scholarly writing demonstrates original ideas, ground-breaking thoughts, and fact-based scientific and technical writing. Four primary traits define educational writing: A strong statement, peer-reviewed resources, an academic style, a distinctive voice, and a consistent structure and citation style are all essential. They employed it to discover novel information in a particular field of study, enabling authors to adopt scientific practices like appropriate research techniques[7]. Scholarly publications are incredibly significant for any academic career, promotion, research program, and many more on a global scale. Writing in scholarly journals and having work published are indicators of a researcher's intellectual productivity[8].

The standard format for research articles is Title, Author, Abstract, Keyword, Introduction, Methods, Results, Discussions, Conclusion, Acknowledgements, References, and Appendices; however, the structures differ from publisher to publisher. The academic article could be an experiment, case study, survey study, or any combination of some other types. It provides first-hand information to the readers[9]. Scholarly writing is an original idea of writing, and thus, it is referred to as the primary source. It is an invaluable resource for researchers. It should be published in reputed and peer-reviewed journal/conference proceedings, where they are archived for a long time as it is used as a reference by many other researchers, information seekers, and academicians[10]. Publishing a research paper is a lengthy and time-consuming process, and it depends on journal periodicity and reviewers of the journals[11].

## 1.2 Importance of Citation

Citations are used to acknowledge authors' original works. It acts as a kind of gratitude and credits, indicating the sources based on which the information has been collected, the resource providers themselves, and the platforms on which respective work has been published and accessed, while also providing details such as the year, volume, and page number. Appropriate citations are essential for assisting information seekers in locating existing research for intellectual objectives and serve as a reward for the original author. To accomplish this, refereed publications must include accurate citations and adhere to the correct style guidelines, which include APA, MLA, Chicago, and many more[12]. Publishers require proper citations in all scholarly articles, conference proceedings, and many other publications; therefore, they utilise various style manual guidelines to achieve this. There are typically two sorts of citations offered: in-text citations and bibliographic citations or references[13]. Sources serve as a liaison between two or more articles[14].

## 1.3 Features of Plagiarism Software

The following are some standard features of plagiarism detection software: it supports uploading files in a variety of multimedia formats, including TXT, HTML, RTF, MS Word, PPTX, XLS, PDF, EPUB, FB2, ODT, and URL; copy-paste text functionality; grammar checking; synonym and vocabulary suggestions; an in-text citation and referencing; sentence re-framing, re-modification, re-arrangement, re-structuring; downloading the plagiarism report; finding the text; and cloud-based, account setup, and document storage, comparisons of specific similarities, support for several languages, the ability to enable web pages, the document's overall score, citation building, submission date, ID, file name, word count, and character count[15-16].

Anti-plagiarism software helps identify textual similarities and for making academicians aware of and able to minimise plagiarism based on its relevance. However, it is not wholly accurate, whereas not all texts are available in the library and websites due to factors like laboratory notes, personal diaries, printed books, and language translation. One linked system in academics' and researchers' lives is plagiarism software[17]. In the journal article screening process, plagiarism software is highly supportable to the editors for checking the similarities and predictability. Also, it supports the standardisation and quality of the journal. It makes the work easier to pre-check the originality of the scientific research manuscript[18].

The most widely used plagiarism detection tools are Grammarly, Turnitin, Urkund, PlagScan, and Unicheck[19]. Paid software offers greater functionality in comparison to free software[20]. Plagiarism is a pervasive problem worldwide, not only among students but also among some academicians. The overwhelming amount of them involves plagiarizing in their writing. Though it cannot totally be controlled, anti-plagiarism software will assist with reducing the number of significant words and phrases. Text that has been plagiarised degrades the quality of the article and tarnishes the reputations of the author, teacher, and institution[21]. Writing that is copied is ubiquitous, and 50% of students agreed. Even though it is a current problem for academic researchers, it is plagiarised. Patch writing involves changing minor words, tweaking the linguistic features, utilizing similar phrases in writing, and technical parroting—which consists in duplicating someone else's work without comprehending it or giving it any thought—are two well-known examples of plagiarism.[22].

## 2. STATEMENT OF THE PROBLEM

University Grants Commission in India accepts Zero Tolerance Policy in research for which they pass the rules and regulations for avoiding plagiarism. In that, they mention that the state and central universities have to check that each of the students. "theses, dissertations, projects, term papers, publications, or any other relevant documents" must be plagiarism free, not only for the students but also for the faculty members[3] (Das, 2018). However, the problem is inadequate funds to subscribe to the plagiarism software.

## 3. OBJECTIVES OF THE STUDY
- To determine whether state university-affiliated colleges have anti-plagiarism software subscriptions.
- To determine whether institutions use anti-plagiarism software to ensure that assignments are original.
- To determine if instructors are educating college students about citation styles and guides.
- To comprehend how people, understand plagiarism detection tools.

## 4. LIMITATIONS OF THE STUDY

Only colleges associated with Tamil Nadu State University were used in this study. Administrators from colleges and librarians participated in the survey since it was deemed appropriate for them.

## 5. METHODOLOGY

Administrators from colleges and library professionals participated in the study. Purposive sampling, a web-based survey approach, and a structured questionnaire were all employed during this study. There are twelve questions. It was divided into two sections. Part one contains four questions regarding the college's details and eight questions concerning plagiarism in part two. Three methods were employed to gather the research data: an official email ID request, which was obtained from the college website; an android application developed by the Delhi Library Association; distribution of the questionnaire hyperlink such as through text messages; and, finally, sharing of the information in WhatsApp groups for Tamil Nadu library professionals. We utilised Google forms to gather the data. Data was gathered between June 9, 2022, and June 16, 2022. 2439 emails in all were sent by the researcher. In this, 2287 emails were delivered, 151 were received, and 152 emails bounced. This study is based on primary data. The study population is Tamil Nadu State University affiliated colleges, and samplings are college administrators and library professionals. There are 22 universities in all, and 18 institutions and their affiliated colleges have responded. Google Sheet was used to import the data, and frequency and percentage were determined.

**Table 1. Information about the respondents' socio-demographics**

| Type | Division | Respondents (%) |
|---|---|---|
| College Type | Government College | 19 (12.6) |
| | Aided College | 22 (14.5) |
| | Self-financing College | 97 (64.2) |
| | Approved Institution | 6 (4) |
| | Constituent College | 6 (4) |
| | Evening Colleges | 1 (0.7) |
| Affiliated University | Alagappa University | 6 (4) |
| | Anna University | 23 (15.2) |
| | Annamalai University | 7 (4.6) |
| | Bharathiar University | 12 (8) |
| | Bharathidasan University | 9 (6) |
| | Madurai Kamaraj University | 7 (4.6) |
| | Manonmaniam Sundaranar University | 4 (2.6) |
| | Mother Teresa Women's University | 2 (1.3) |
| | Periyar University | 12 (8) |
| | Tamil Nadu Agricultural University | 3 (2) |
| | Tamil Nadu Dr. J. Jayalalithaa Fisheries University | 1 (0.7) |
| | Tamil Nadu Open University | 0 (0.0) |
| | Tamil Nadu Physical Education and Sports University | 2 (1.3) |
| | Tamil Nadu Teachers Education University | 10 (6.6) |
| | Tamil Nadu Veterinary and Animal Sciences University | 0 (0.0) |
| | Tamil University | 0 (0.0) |
| | The Tamilnadu Dr. Ambedkar Law University | 3 (2) |
| | The Tamil Nadu Dr. J. Jayalalitha Music and Fine Arts University | 2 (1.3) |
| | The Tamil Nadu Dr. M.G.R. Medical University | 13 (8.6) |
| | Tamil Nadu National Law University | 0 (0.0) |
| | Thiruvalluvar University | 14 (9.3) |
| | University of Madras | 21 (13.9) |
| Designation | Principal | 24 (15.9) |
| | Library Professional | 97 (64.2) |
| | College Administrators (Registrar, Director, COE Etc) | 30 (19.9) |
| Courses offered (Multiple options allowed (N=151)) | UG (Undergraduate) | 146 (96.7) |
| | PG (Postgraduate) | 115 (76.2) |
| | M.Phil (Master of Philosophy) | 38 (25.2) |
| | PhD (Doctor of Philosophy) | 40 (26.5) |
| | Diploma | 20 (13.2) |
| | Certificate | 12 (8) |

## 6. DATA ANALYSIS AND INTERPRETATION

The respondents' socio-demographic information is displayed in Table 1. Most responders, in terms of college type, are from self-financing institutions. More respondents come from Anna University, followed by Madras University, from the list of colleges affiliated with a state university. Professionals in the library industry make up a significant percentage of the respondents. Multiple answers were permitted for these questions in the courses that colleges offered. It demonstrates that 76.2 % of colleges offer postgraduate courses, with 96.7 % of colleges offering undergraduate courses.

Whether or not colleges have anti-plagiarism software is shown in Table 2. The results indicate that 70.9 % did not subscribe, while 29.1 per cent were subscribers.

Table 2. Status of employing anti-plagiarism software

| Plagiarism software subscribed | Respondents (%) |
|---|---|
| Yes | 44 (29.1) |
| No | 107 (70.9) |

Table 3. Name of the acquired anti-plagiarism tool

| Software tool | Respondents (%) |
|---|---|
| Grammarly | 7 (4.6) |
| Ouriginal (Urkund) | 23 (15.2) |
| Turnitin | 14 (9.3) |
| Not subscribed | 107 (70.9) |

The subscribed plagiarism software details are displayed in Table 3. The findings indicate that 70.2 per cent of respondents used no anti-plagiarism software, while the remaining 15.2 per cent utilised Ouriginal (Urkund), followed by Turnitin, 10 per cent, and Grammarly, 4.6 per cent.

The justifications for not using plagiarism detection software are listed in Table 4. 30.5 per cent of respondents declared they didn't need it. In comparison, 41.7 per cent reported they didn't subscribe because it was too expensive. Over a quarter of respondents—27 per cent—said they could not subscribe due to inadequate funding, 23.2 per cent said they intended to do so during the following academic year, and 8.6 per cent claimed it was unreliable.

Table 4. Justification for not using plagiarism detection tools

| Reasons for not subscribed | Respondents (N=151) (%) |
|---|---|
| For my college, it is not necessary | 46 (30.5) |
| Too expensive | 63 (41.7) |
| Lack of budget allocation | 42 (27.8) |
| Non-reliable | 13 (8.6) |
| Planning to subscribe in the next academic year | 35 (23.2) |

According to Table 5, 33.1 % of colleges indicated that they had never given pupils original assignments and projects, whereas more than one-fourth of 27.2 per cent of respondents were unsure. 39.7 % of colleges acknowledged giving the students original projects or assignments.

Table 5. Whether college offers original projects and assignments to its pupils

| College gives plagiarism-free assignments/projects to the students | Respondents (%) |
|---|---|
| Yes | 60 (39.7) |
| No | 50 (33.1) |
| Not Sure | 41 (27.2) |

According to Table 6, 19.9 % of respondents' colleges use anti-plagiarism software to examine student projects and assignments for originality. Similarly, 19.9 % of respondents employ free online plagiarism detection tools, and in the vast majority of colleges, 60.2 % are uncertain.

Table 6. Technique to ensure that students' projects and assignments are original

| Methods followed to identify the plagiarism | Respondents (%) |
|---|---|
| Subscribed anti-plagiarism software | 30 (19.9) |
| Free online anti-plagiarism tools | 30 (19.9) |
| Not sure | 91 (60.2) |

With the need to enhance student and teacher awareness about plagiarism, frequent training and awareness programs are needed, so the same was asked of these participants. Table 7 indicates that only 29.1 per cent of colleges are conducting training and awareness programs about plagiarism,

Table 7. Any training sessions that the college gave to raise awareness of plagiarism

| The college organised an awareness program on plagiarism | Respondents (%) |
|---|---|
| Yes | 44 (29.1) |
| No | 107 (70.9) |

but unfortunately, the remaining 70.9 per cent are not conducting such programs.

Table 8 illustrates that the majority of respondents, 88.1 per cent, believe that plagiarism software is used to identify plagiarised content, whereas 37.8 per cent of respondents believe plagiarism software does not detect written materials. Almost a quarter of respondents, 24.5 per cent, feel plagiarism software fails to detect scientific symbols, and 12 per cent of respondents believe plagiarism software fails to detect citations.

Table 8. Perceptions of plagiarism detection software

| Plagiarism software perception | Respondents (N=151) (%) |
| --- | --- |
| To identify the plagiarized content | 133 (88.1) |
| Fails to detect the citations | 18 (12) |
| Fails to detect scientific symbols | 37 (24.5) |
| Fails to detect written materials | 57 (37.8) |

Note: allowed to select more than one option, so the percentage is more than 100

## 7. DISCUSSION

Plagiarism is widely happening, but it is against academic ethics. Plagiarism's main reason is not giving a proper citation to the original resources; instead, they provide unauthorised or duplicate resources. For instance, a surprisingly large number of people utilise search engines to seek information, which displays several websites. Information seekers believed they were providing an accurate source. Still, in practice, the top piece of information from search results is secondary information that has been copied by website and weblog creators. As a result, to conduct their studies properly, researchers occasionally need to use plagiarism detection software to locate authentic sources[23].

According to a survey, most colleges are reluctant to employ plagiarism software because it is too expensive, there are financial constraints, or it is not necessary. However, most colleges plan to use plagiarism software in the upcoming year. By the survey report, 30.5 per cent of respondents felt obtaining a license for plagiarism detection software was unnecessary. This result reveals the respondent's lack of understanding of plagiarism and detection software. They are not able to check the plagiarism of their faculty members and students without plagiarism software. Publications are essential for promoting the college and strengthening its academic reputation[24]. To control plagiarism, before submitting the paper, the academicians and authors have to check with any plagiarism tools; higher education institutes have to subscribe to plagiarism software because students cannot afford to buy the software as well as each student should be given a submission assignment, project and thesis to check-in plagiarism tool; librarians and publishers have to take prime responsibility of controlling the plagiarism. A lack of creative writing and critical thinking is the reason for plagiarism. Among academicians, limited knowledge and awareness is another reason for plagiarism[25-26]. India is a multilingual nation with few native speakers of English. Low academic literacy and citation levels are other reasons for plagiarism and inadequate language[27]. The study report found that 70.9 per cent of respondents' colleges are not subscribed to plagiarism software, as it is too expensive. Overall, the study demonstrates the need for subscribing to plagiarism software by all the college and university libraries and explicitly mentions the significance of the library website.

## 8. SUGGESTION

This study recommended that colleges try to subscribe to plagiarism detection Software. University Grants Commission should indeed, at last, step up to the plate and will provide the free plagiarism detection software tool for all universities and colleges rather than paying because some colleges cannot subscribe because it is costly. University Grants Commission should always be given monetary support for subscribing to such research-oriented software. Colleges must frequently organise programs to educate both students and staff about plagiarism. They must also strictly enforce the requirement that all college projects be original and free of plagiarism.

## 9. CONCLUSION

UGC takes great care to uphold academic integrity, and the Indian Government actively encourages excellent research. The majority of institutions, however, do not use technologies to identify plagiarism in their research programs. This study unequivocally shows that plagiarism detection software is required for all colleges because, under UGC guidelines, completing a PhD requires at least two publications. Checking one's thesis and dissertation for plagiarism software is crucial, and submitting a research paper before doing so benefits everyone. The software will assist in identifying the similarities in these kinds of circumstances. API (Academic Performance Indicator) scores are strictly followed for research and academic contributions based on the teacher's evaluation of their accomplishment. It is divided into three categories—the third category is specifically for intellectual contributions and research. Various degrees of advancement in universities and colleges have different minimum API score requirements for teachers in this category. The maximum score for publication is 60/yr minimum varies from one post to another rank. Based on UGC guidelines, appointment and promotions API score is calculated in this publication is an essential role. Without software, faculties are unable to check their research paper's similarities. So that is a reason plagiarism software is necessary for all colleges.

This survey examines the status of the institutions and whether they are seriously using the plagiarism detection software or not to check the scholarly content, and this study will eye-opener some stakeholders to know about the status of plagiarism software. We used Tamil Nadu state colleges as our sample because it is one of the states in India that produces the most PhDs and research papers[28]. As a result, our study will undoubtedly benefit not only the stakeholders in Tamil Nadu but the entire nation as well.

## ACKNOWLEDGEMENT

We would like to express our sincere appreciation to our advisor Professor Dr R. Sevukan, Department of Library and Information Science, Pondicherry University, for his assistance with this research study. The help provided by Dr. Manuel Raj Peter, Associate Director (Public Services, General Admin sand Collection Development) at the Jio Institute in Mumbai, Maharashtra, is much appreciated. We are grateful to the members of the Library Association of India who contributed to this research by taking the time to answer this survey question. Not least, the reviewers' remarks help make this paper shine. Thank you for the insightful advice.

## CONTRIBUTORS

**Mr Subaveerapandiyan A.** is working as a University Librarian in DMI-St. Eugene University, Lusaka, Zambia. His areas of interest include: Digital literacy, scholarly communication, knowledge management, research data management, open access and publishing.
In this study, his contribution is to research design, preparing research questionnaires, e-mail IDs collection from respective university websites, data collection, data analysis and tabulation.

**Mr Sakthivel N.** is a Library Intern at the Jio Institute Digital Library, Jio Institute, Mumbai, India. He has completed his MLibISc in Pondicherry University. His areas of interest include: Digital library, information seeking behaviour, research data management and information literacy.
In this study, his contribution is collection of e-mail IDs from respective university websites, data collection, data interpretation, discussion and conclusion.